\documentclass[12pt,a4paper]{article}
\usepackage[a4paper,margin=1in]{geometry}
\usepackage{amsmath,amssymb,bm}
\usepackage{graphicx}
\usepackage{cite}
\usepackage[colorlinks=true,linkcolor=blue,citecolor=blue,urlcolor=blue]{hyperref}
\usepackage{authblk}
\usepackage{setspace}
\title{Relativistic saturation of Coulomb-induced electron decoherence:\\
from eikonal phase noise to Bethe--Salpeter kinetic theory}

\author[1,2]{Yury A. Budkov\thanks{Email: \texttt{ybudkov@hse.ru}}}
\affil[1]{Laboratory of Computational Physics, HSE University, Tallinskaya st. 34, 123458 Moscow, Russia}
\affil[2]{Frumkin Institute of Physical Chemistry and Electrochemistry, Russian Academy of Sciences, 31-4 Leninsky Prospect, 119071 Moscow, Russia}
\date{}

\begin{document}

\maketitle

\begin{abstract}
We develop a microscopic theory of the mutual coherence and small-angle scattering of relativistic electron matter waves propagating through a Coulomb-fluctuating medium. 
Starting from the Dirac equation, we derive a relativistic paraxial wave equation and then obtain, rather than assume, the Bethe--Salpeter and Wigner--Boltzmann kinetic equations for the disorder-averaged two-point coherence. The Coulomb environment enters through the eikonal phase-coupling factor $A_{\rm rel}=1/(\hbar v)$, which saturates at $1/(\hbar c)$ for ultra-relativistic particles. This saturation is the central physical result: increasing the beam voltage suppresses Coulomb phase noise only up to a finite relativistic floor. We first formulate the eikonal approximation, in which the transmitted electron wave accumulates a random longitudinal phase. For a one-component Coulomb medium the electrostatic potential correlator retains a Coulomb tail, producing a logarithmic phase structure function and an algebraic decay of the transverse mutual coherence. We then go beyond the eikonal construction by deriving the Bethe--Salpeter equation for the Cooperon-like two-point coherence propagator directly from the paraxial wave equation. After a longitudinal Markov projection this equation reduces, after a Wigner transform, to a kinetic equation with a Coulomb small-angle scattering kernel. The coordinate-space solution of the same Bethe--Salpeter equation recovers the eikonal mutual coherence function, while the Wigner form additionally describes angular diffusion and transverse broadening caused by multiple small-angle scattering. The resulting theory separates three effects which are often conflated in charged-particle wave propagation: attenuation of the coherent amplitude, decay of mutual coherence, and conservative redistribution of intensity in transverse phase space. The results provide a microscopic framework for assessing medium-induced limits to electron-wave coherence in liquid-cell and cryogenic electron microscopy and show why increasing the beam energy eventually enters a regime of diminishing returns for suppressing Coulomb-disorder-induced decoherence.
\end{abstract}


\vspace{1em}

\section{Introduction}
\label{sec:intro}

The propagation of coherent waves through random media is a common theme in condensed-matter physics, optics, plasma physics, and electron microscopy. 
For quantum particles, disorder may suppress transport through Anderson localization~\cite{Anderson1958,LeeRamakrishnan1985,KramerMacKinnon1993}. 
For classical and quantum waves, the same microscopic physics can also be described in terms of ladder diagrams, Bethe--Salpeter equations, Wigner functions, and radiative-transfer or kinetic equations~\cite{Foldy1945,Lax1951,Ishimaru1978,VanRossumNieuwenhuizen1999,AkkermansMontambaux2007,RyzhikPapanicolaouKeller1996}. 
The Wigner-transform route to kinetic transport is well established for waves in random media. 
Ryzhik, Papanicolaou, and Keller derived transport equations for wave energy in random media~\cite{RyzhikPapanicolaouKeller1996}, while Bal, Komorowski, and Ryzhik established self-averaging properties of Wigner transforms in the high-frequency random-medium limit~\cite{BalKomorowskiRyzhik2003}. 
For paraxial propagation, Fannjiang developed two-frequency Wigner and radiative-transfer descriptions in which spatial spread, coherence length, and coherence bandwidth are obtained from Boltzmann- or Fokker--Planck-type phase-space equations~\cite{Fannjiang2007TwoFrequencyRT}. 
Closely related white-noise paraxial models were studied by Garnier and Solna, who obtained closed transport equations for Wigner distributions and field autocorrelations~\cite{GarnierSolna2009}. 
These works provide the methodological background for the kinetic part of the present paper. 
Our use of the Wigner representation is therefore not meant as a new transport formalism by itself; the new element is the application to the mutual coherence of a relativistic electron matter wave in a Coulomb-fluctuating medium, with a disorder spectrum fixed by equilibrium electrostatics and with a phase-coupling constant that saturates in the relativistic limit. 
The different languages emphasize complementary observables. 
The single-particle Green function measures the survival of a coherent amplitude; the two-particle propagator measures mutual coherence and intensity transport; the Wigner function resolves the redistribution of intensity in transverse phase space.

From the point of view of charged-particle wave physics, the Coulomb medium acts simultaneously as a random phase grating and as a source of conservative small-angle scattering. The question addressed below is therefore not only how much of the coherent amplitude survives disorder averaging, but also how the mutual coherence of an electron matter wave is degraded and how the same microscopic Coulomb fluctuations redistribute the transverse wave-vector content of the beam.

In electron microscopy of liquids and soft matter, this distinction is practically important. 
Modern liquid-cell transmission electron microscopy (TEM) and cryogenic electron microscopy are usually discussed in terms of instrumental aberrations, radiation damage, inelastic scattering, sample thickness, and stability~\cite{Frank2006,DeJonge2019,Ross2015}. 
However, a liquid or hydrated specimen also contains thermally fluctuating charges. 
Even if the average medium is electrically neutral and Debye screened, the instantaneous electrostatic potential seen by a fast electron is random. 
The accumulated phase noise generated by this Coulomb disorder provides a microscopic mechanism of mutual-coherence degradation which is not removed by improving the electron optics. 
It is therefore useful to ask whether an electron beam has a Coulomb-limited coherence scale, in close analogy with wave-front degradation in statistical optics~\cite{Tatarskii1961,Rytov1989,Goodman2015,MandelWolf1995}.

The present work is motivated by two simple observations. 
First, in a model Coulomb medium, represented for definiteness by a classical one-component plasma (OCP), the correlator of the electrostatic potential generated by equilibrium density fluctuations has a long-distance Coulomb tail. 
Although charge-density correlations are screened on the Debye length, the potential correlator behaves as $1/r$ at large $r$. 
This is a standard consequence of Coulomb statistical mechanics~\cite{BausHansen1980,Ichimaru1982}. 
Second, a high-energy electron in a TEM is not non-relativistic. 
A quantitatively useful theory must begin with the relativistic paraxial reduction of the Dirac equation and identify the correct coupling of the electron envelope to the electrostatic disorder.

The key result of this reduction is the effective coupling constant
\begin{equation}
A_{\rm rel}=\frac{1}{\hbar v}=\frac{1}{\hbar c\beta}.
\label{eq:Arel_intro}
\end{equation}
This factor is exact to leading paraxial order in the scalar potential. It has the usual non-relativistic eikonal form and, for $v\to c$, saturates:
\begin{equation}
\lim_{v\to c}A_{\rm rel}=\frac{1}{\hbar c}.
\label{eq:Arel_saturation_intro}
\end{equation}
Thus, relativistic acceleration reduces the Coulomb phase noise only up to a finite bound. 
This is the main physical message of the paper. 
The electron wavelength and lens aberrations may continue to improve with increasing beam energy, but the coupling of the transmitted wave front to thermally fluctuating Coulomb fields has an irreducible high-energy floor.

The theory is developed in two stages. 
First, we present the straight-line eikonal approximation. 
It gives a transparent expression for the accumulated phase, the phase structure function, the algebraic long-distance decay of the mutual coherence, and the relativistically saturated attenuation scale. 
Second, we derive the Bethe--Salpeter equation for the two-point coherence function from the full paraxial equation. 
After a Wigner transform this equation becomes a kinetic equation with a Coulomb small-angle scattering kernel. 
This derivation shows that the eikonal result is the homogeneous-coordinate limit of a more general paraxial transport theory. 
The Wigner equation also makes clear what is meant by multiple scattering in the present problem: intensity is redistributed in transverse momentum, whereas mutual coherence decays through the same Coulomb-induced two-point dephasing kernel.

The paper is organized as follows. 
Section~\ref{sec:model} introduces the Coulomb disorder model and the relativistic paraxial equation. 
Section~\ref{sec:eikonal} derives the eikonal coherence function and the relativistic saturation of the coupling, attenuation scale, and coherence length. 
Section~\ref{sec:bs} derives the Bethe--Salpeter equation for the Cooperon-like coherence propagator. 
Section~\ref{sec:wigner} performs the Wigner transform and obtains the kinetic equation. 
Section~\ref{sec:coulomb_kernel} evaluates the Coulomb kernel and shows how the eikonal algebraic coherence tail is recovered from the Bethe--Salpeter formulation. 
Section~\ref{sec:implications} discusses physical consequences for electron microscopy, and Sec.~\ref{sec:conclusion} summarizes the results.

\section{Coulomb disorder and relativistic paraxial equation}
\label{sec:model}

We consider a quantum test electron propagating predominantly along the $z$ direction through a classical Coulomb-fluctuating medium. 
The transverse coordinate is denoted by $\bm \rho=(x,y)$. 
The random electrostatic potential energy acting on the electron is denoted by $W(\bm \rho,z)$. 
Throughout the paper the disorder is assumed to be statistically homogeneous, Gaussian, and static during the electron transit time. 
The static approximation is appropriate when the transit time through the sample is shorter than the characteristic ionic relaxation time. 
For TEM velocities this condition is often less restrictive than the corresponding condition for slow probe particles.

For an OCP of ions of charge $q$ and mean density $n_0$ in a neutralizing background, the random potential acting on a test charge $q_0$ is
\begin{equation}
W(\mathbf r)=q_0\int d^3r'\,
\frac{q\,\delta n(\mathbf r')}{\varepsilon|\mathbf r-\mathbf r'|},
\label{eq:W_def}
\end{equation}
where $\delta n$ is the ionic density fluctuation and $\varepsilon$ is the dielectric permittivity. 
Within the random-phase approximation the potential is a Gaussian field with zero mean and correlator~\cite{budkov2024statistical} (see also Appendix \ref{app:ocp_correlator})
\begin{equation}
K(r)\equiv
\langle W(\mathbf r)W(0)\rangle
=
C\,\frac{1-e^{-\kappa r}}{r},
\label{eq:K_real}
\end{equation}
where
\begin{equation}
C=\frac{k_B T q_0^2}{\varepsilon},
\qquad
\kappa=\lambda_D^{-1}.
\label{eq:C_kappa}
\end{equation}
The three-dimensional Fourier transform is
\begin{equation}
\widetilde K_3(\mathbf p)
=
4\pi C
\left[
\frac{1}{p^2}
-
\frac{1}{p^2+\kappa^2}
\right]
=
\frac{4\pi C\kappa^2}{p^2(p^2+\kappa^2)}.
\label{eq:K3}
\end{equation}
The small-$p$ limit is $\widetilde K_3(\mathbf p)\simeq4\pi C/p^2$, which is the Fourier representation of the long-range Coulomb tail of the potential correlator.

The electron is described by a slowly varying paraxial envelope. 
Starting from the stationary Dirac equation in a weak scalar electrostatic potential and eliminating the small spinor component to leading paraxial order gives an effective scalar equation for the upper envelope (see Appendix \ref{app:dirac}),
\begin{equation}
i\partial_z\psi(\bm\rho,z)
=
\left[
-\frac{1}{2k}\nabla_\perp^2
+
A_{\rm rel}W(\bm\rho,z)
\right]\psi(\bm\rho,z).
\label{eq:paraxial}
\end{equation}
Here
\begin{equation}
k=\frac{\gamma m v}{\hbar},
\qquad
A_{\rm rel}=\frac{1}{\hbar v}=\frac{1}{\hbar c\beta}.
\label{eq:k_A}
\end{equation}
Spin-dependent Darwin and spin-orbit corrections are of higher paraxial order and will not be considered here. 
Equation~\eqref{eq:paraxial} is the central microscopic wave equation for the transmitted electron matter wave. 
The variable $z$ plays the role of time, the transverse kinetic term describes diffraction, and the Coulomb disorder potential produces phase modulation and small-angle scattering.

The corresponding limiting forms are
\begin{equation}
A_{\rm rel}=\frac{1}{\hbar v},
\qquad
v\ll c,
\label{eq:A_NR}
\end{equation}
and
\begin{equation}
A_{\rm rel}\to\frac{1}{\hbar c},
\qquad
\gamma\to\infty.
\label{eq:A_UR}
\end{equation}
Thus the relativistic effect is not an additional spinor prefactor multiplying the eikonal phase. Rather, it is the saturation of the longitudinal velocity at $c$, which prevents the Coulomb phase coupling from decreasing below the finite value~\eqref{eq:A_UR}.

\section{Eikonal approximation and mutual coherence}
\label{sec:eikonal}

We first recall the eikonal theory because it provides the simplest physical interpretation of the decoherence mechanism. 
The eikonal approximation neglects transverse diffraction during propagation through the sample. 
The electron then follows a straight line at fixed transverse coordinate $\bm\rho$ and accumulates the random phase
\begin{equation}
\phi(\bm\rho)
=
-A_{\rm rel}\int_0^L dz\,W(\bm\rho,z).
\label{eq:eikonal_phase}
\end{equation}
The mutual coherence between two points separated by $\bm\rho$ on the exit plane is
\begin{equation}
\gamma(\bm\rho)
=
\left\langle
\exp\{i[\phi(\bm\rho)-\phi(0)]\}
\right\rangle.
\label{eq:gamma_def}
\end{equation}
Since $W$ is Gaussian, the second cumulant is exact:
\begin{equation}
\gamma(\rho)
=
\exp\left[-\frac{1}{2}D_\phi(\rho)\right],
\label{eq:gamma_cumulant}
\end{equation}
where the phase structure function is
\begin{equation}
D_\phi(\rho)
=
\left\langle
[\phi(\bm\rho)-\phi(0)]^2
\right\rangle.
\label{eq:D_def}
\end{equation}
Substitution of Eq.~\eqref{eq:eikonal_phase} gives
\begin{equation}
D_\phi(\rho)
=
2A_{\rm rel}^2
\int_0^L dz\int_0^L dz'\,
\left[
K(|z-z'|)
-
K\!\left(\sqrt{\rho^2+(z-z')^2}\right)
\right].
\label{eq:D_double}
\end{equation}
For $L$ large compared with the Debye length, and for transverse separations not too close to the sample thickness, the leading term is
\begin{equation}
D_\phi(\rho)
=
4A_{\rm rel}^2LC
\int_0^{\kappa\rho}dt\,
\left[
\frac{1}{t}-K_1(t)
\right],
\label{eq:D_exact}
\end{equation}
where $K_1$ is a modified Bessel function. 
Using $d[\ln t+K_0(t)]/dt=1/t-K_1(t)$, this can also be written as
\begin{equation}
D_\phi(\rho)
=
4A_{\rm rel}^2LC
\left[
\ln(\kappa\rho)+K_0(\kappa\rho)+\gamma_E-\ln2
\right].
\label{eq:D_K0}
\end{equation}
The small-separation limit $\kappa\rho\ll1$ is
\begin{equation}
D_\phi(\rho)
\simeq
A_{\rm rel}^2LC\,(\kappa\rho)^2
\left[
\ln\frac{2}{\kappa\rho}+1-\gamma_E
\right],
\label{eq:D_small}
\end{equation}
whereas at large separations $\kappa\rho\gg1$,
\begin{equation}
D_\phi(\rho)
\simeq
4A_{\rm rel}^2LC
\left[
\ln(\kappa\rho)+\gamma_E-\ln2
\right].
\label{eq:D_large}
\end{equation}
Thus the mutual coherence has an algebraic tail:
\begin{equation}
\gamma(\rho)\propto(\kappa\rho)^{-\eta_{\rm rel}},
\qquad
\eta_{\rm rel}=2A_{\rm rel}^2LC.
\label{eq:algebraic}
\end{equation}
This is the main eikonal result. 
It does not mean that the microscopic charge correlations are unscreened. 
Rather, the longitudinal projection of the Coulomb potential correlator produces a logarithmic transverse phase structure function.

A convenient operational coherence length $\rho_c$ may be defined by $D_\phi(\rho_c)=2$. 
Using the small-distance form~\eqref{eq:D_small} gives the scaling estimate
\begin{equation}
\rho_c
\sim
\lambda_D
\sqrt{\frac{\ell_{\rm rel}}{L}},
\label{eq:rho_c}
\end{equation}
where the corresponding coherent-amplitude attenuation scale is
\begin{equation}
\ell_{\rm rel}
=
\frac{1}{A_{\rm rel}^2C\ln(\kappa L_s)}.
\label{eq:ell_rel}
\end{equation}
Here $L_s$ is the longitudinal infrared scale entering the Coulomb logarithm. 
Substituting Eq.~\eqref{eq:k_A}, we obtain
\begin{equation}
\ell_{\rm rel}
=
\frac{\hbar^2 v^2}{C\ln(\kappa L_s)}.
\label{eq:ell_explicit}
\end{equation}
In the ultra-relativistic limit,
\begin{equation}
\ell_{\rm rel}
\to
\ell_{\rm max}
=
\frac{\hbar^2c^2}{C\ln(\kappa L_s)}.
\label{eq:ell_max}
\end{equation}
Consequently the coherence length also saturates:
\begin{equation}
\rho_c
\lesssim
\lambda_D
\sqrt{\frac{\ell_{\rm max}}{L}}.
\label{eq:rho_sat}
\end{equation}
This saturation is the central experimental message. 
A higher accelerating voltage reduces Coulomb decoherence only until the relativistic floor in $A_{\rm rel}$ is reached.

The eikonal theory is transparent, but it also has a limitation: it suppresses transverse diffraction and multiple small-angle scattering. 
To show that the coherence result is not merely an artifact of straight-line propagation, we now derive the Bethe--Salpeter equation from the full paraxial wave equation.

\section{Bethe--Salpeter equation for the coherence propagator}
\label{sec:bs}

For a fixed realization of the disorder define the two-point coherence matrix
\begin{equation}
J_W(\bm\rho_1,\bm\rho_2,z)
=
\psi_W(\bm\rho_1,z)\psi_W^*(\bm\rho_2,z).
\label{eq:J_def}
\end{equation}
Using Eq.~\eqref{eq:paraxial} and its complex conjugate, one obtains the exact equation
\begin{equation}
\partial_zJ_W
=
\frac{i}{2k}
(\nabla_1^2-\nabla_2^2)J_W
-
iA_{\rm rel}
[W(\bm\rho_1,z)-W(\bm\rho_2,z)]J_W.
\label{eq:J_eq}
\end{equation}
The disorder-averaged two-particle object
\begin{equation}
\Gamma(\bm\rho_1,\bm\rho_2,z)
=
\langle J_W(\bm\rho_1,\bm\rho_2,z)\rangle_W
\label{eq:Gamma_def}
\end{equation}
is the paraxial Cooperon-like coherence propagator. 
It is the retarded-advanced ladder object appropriate for mutual coherence.

To close the equation we use the Markov approximation along the propagation direction:
\begin{equation}
\langle
W(\bm\rho,z)W(\bm\rho',z')
\rangle
=
\delta(z-z')B(\bm\rho-\bm\rho').
\label{eq:Markov}
\end{equation}
The Markov approximation in Eq.~\eqref{eq:Markov} should be understood as a longitudinal projection of the static three-dimensional correlator onto the paraxial propagation direction. For a short-range random medium the longitudinal correlation length would be microscopic. In the present Coulomb problem, however, the potential correlator has a long-range tail, and the longitudinal correlation length of a transverse Fourier component $Q$ is of order $Q^{-1}$ for $Q\ll\kappa$. Thus the Markov reduction is controlled for transverse modes satisfying $QL\gg1$, while the Debye length provides the ultraviolet cutoff $Q\lesssim\kappa$. Equivalently, the leading-logarithmic regime is the window $L^{-1}\ll Q\ll\kappa$. The modes near the lower cutoff $Q\sim L^{-1}$ are not strictly local in $z$ and can change only non-universal constants under the logarithm. The leading Coulomb logarithm and the kernel $\widetilde B(Q)=\widetilde K_3(Q,0)$ are therefore retained within logarithmic accuracy. This longitudinal Markov projection is distinct from the frozen-disorder approximation, which requires the electron transit time $L/v$ to be shorter than the characteristic ionic relaxation time.
The transverse correlator $B$ is the longitudinal integral of the three-dimensional correlator,
\begin{equation}
B(\bm\rho)
=
\int_{-\infty}^{\infty}du\,
K\!\left(\sqrt{\rho^2+u^2}\right).
\label{eq:B_real}
\end{equation}
Equivalently, in two-dimensional Fourier space,
\begin{equation}
\widetilde B(\mathbf Q)
=
\widetilde K_3(\mathbf Q,q_z=0).
\label{eq:B_tilde_def}
\end{equation}
Within this effective longitudinal Markov representation, which preserves the integrated transverse kernel $B(\bm\rho)$, the Gaussian average over the last term in Eq.~\eqref{eq:J_eq} gives
\begin{equation}
\partial_z\Gamma
=
\frac{i}{2k}
(\nabla_1^2-\nabla_2^2)\Gamma
-
A_{\rm rel}^2
[
B(0)-B(\bm\rho_1-\bm\rho_2)
]\Gamma.
\label{eq:BS_rho}
\end{equation}
This is the Bethe--Salpeter equation~\cite{BetheSalpeter1951} in its paraxial Markov form. 
The term $B(0)-B(\bm\rho_1-\bm\rho_2)$ is the dephasing rate of two paths separated by $\bm\rho_1-\bm\rho_2$. 
Common phase fluctuations cancel; only differential fluctuations destroy mutual coherence.

Note that for a Gaussian random potential the closure of Eq.~(\ref{eq:BS_rho})
follows from the Novikov identity~\cite{Novikov1965},
\[
\langle W({\boldsymbol\rho},z)F[W]\rangle
=
\int dz'\,d^2\rho'\,
\langle W({\boldsymbol\rho},z)W({\boldsymbol\rho}',z')\rangle
\left\langle
\frac{\delta F[W]}{\delta W({\boldsymbol\rho}',z')}
\right\rangle .
\]
With the longitudinal Markov correlator
\(\langle W({\boldsymbol\rho},z)W({\boldsymbol\rho}',z')\rangle
=\delta(z-z')B({\boldsymbol\rho}-{\boldsymbol\rho}')\),
this relation gives the local dephasing kernel
\(A_{\rm rel}^{2}[B(0)-B({\boldsymbol\rho}_{1}-{\boldsymbol\rho}_{2})]\).

It is useful to introduce center and relative coordinates,
\begin{equation}
\mathbf R=\frac{\bm\rho_1+\bm\rho_2}{2},
\qquad
\mathbf r=\bm\rho_1-\bm\rho_2.
\label{eq:R_r}
\end{equation}
Since
\begin{equation}
\nabla_1^2-\nabla_2^2=2\nabla_R\cdot\nabla_r,
\label{eq:lap_diff}
\end{equation}
Eq.~\eqref{eq:BS_rho} becomes
\begin{equation}
\partial_z\Gamma(\mathbf R,\mathbf r,z)
=
\frac{i}{k}
\nabla_R\cdot\nabla_r\,\Gamma
-
A_{\rm rel}^2
[B(0)-B(\mathbf r)]\Gamma.
\label{eq:BS_Rr}
\end{equation}
The first term describes paraxial diffraction and transverse transport of correlations. 
The second term describes Coulomb dephasing. 
The eikonal limit corresponds to omitting the first term, or to considering a spatially homogeneous beam for which $\nabla_R\Gamma=0$.

\section{Wigner kinetic equation}
\label{sec:wigner}

The Wigner representation~\cite{Wigner1932,Bastiaans1986} converts the Bethe--Salpeter equation into a transverse kinetic equation. 
This step follows the standard logic of Wigner and radiative-transfer limits for random waves~\cite{RyzhikPapanicolaouKeller1996,BalKomorowskiRyzhik2003,Fannjiang2007TwoFrequencyRT,GarnierSolna2009}. 
The kinetic equation is not postulated phenomenologically. It is obtained by starting from the Dirac equation for a relativistic electron in a weak scalar potential, performing the paraxial reduction, constructing the disorder-averaged two-point coherence propagator, and applying the Wigner transform to the resulting Bethe--Salpeter equation. The Coulomb collision kernel is therefore fixed by the equilibrium electrostatic potential correlator of the medium, rather than introduced as an adjustable ansatz. 
Two specializations are worth emphasizing. 
First, the object transformed here is the Cooperon-like mutual-coherence propagator of a charged quantum wave, rather than only an energy density of a classical scalar wave. 
Second, the scattering kernel contains the same relativistically saturated coupling $A_{\rm rel}$ as the eikonal phase theory. 
Define
\begin{equation}
I(\mathbf R,\mathbf p,z)
=
\int d^2r\,
e^{-i\mathbf p\cdot\mathbf r}
\Gamma(\mathbf R,\mathbf r,z).
\label{eq:Wigner_def}
\end{equation}
The inverse transform is
\begin{equation}
\Gamma(\mathbf R,\mathbf r,z)
=
\int\frac{d^2p}{(2\pi)^2}
e^{i\mathbf p\cdot\mathbf r}
I(\mathbf R,\mathbf p,z).
\label{eq:Wigner_inv}
\end{equation}
The free term transforms as
\begin{equation}
\int d^2r\,e^{-i\mathbf p\cdot\mathbf r}
\frac{i}{k}\nabla_R\cdot\nabla_r\Gamma
=
-\frac{\mathbf p}{k}\cdot\nabla_R I.
\label{eq:free_Wigner}
\end{equation}
For the dephasing term we write
\begin{equation}
B(\mathbf r)
=
\int\frac{d^2Q}{(2\pi)^2}
e^{i\mathbf Q\cdot\mathbf r}\widetilde B(\mathbf Q).
\label{eq:B_fourier}
\end{equation}
Using
\begin{equation}
B(0)=
\int\frac{d^2Q}{(2\pi)^2}\widetilde B(\mathbf Q),
\label{eq:B0}
\end{equation}
we find
\begin{align}
\partial_z I(\mathbf R,\mathbf p,z)
+
\frac{\mathbf p}{k}\cdot\nabla_R I(\mathbf R,\mathbf p,z)
&=
A_{\rm rel}^2
\int\frac{d^2Q}{(2\pi)^2}
\widetilde B(Q)
\left[
I(\mathbf R,\mathbf p-\mathbf Q,z)
-
I(\mathbf R,\mathbf p,z)
\right].
\label{eq:Wigner_kinetic}
\end{align}
This is the Wigner kinetic equation for the paraxial Cooperon. 
It has the same structure as a Boltzmann equation for small-angle scattering, as in the radiative-transfer limit of random-wave theory, but here the kinetic density is the phase-space representation of the two-point coherence function. 
This distinction is important: the equation conserves total intensity in phase space while the coordinate-space off-diagonal coherence may decay. 
The collision kernel is
\begin{equation}
w(Q)=A_{\rm rel}^2\widetilde B(Q).
\label{eq:w_kernel}
\end{equation}
The first term on the right-hand side is the gain term, while the second is the loss term. 
The total Wigner intensity is conserved by the collision operator:
\begin{equation}
\int d^2R\int\frac{d^2p}{(2\pi)^2}\,
I(\mathbf R,\mathbf p,z)
=
{\rm const}.
\label{eq:intensity_conservation}
\end{equation}
Thus the kinetic equation does not describe absorption. 
It describes conservative redistribution of intensity in transverse momentum and position, while the off-diagonal coherence in coordinate space decays.

\section{Coulomb scattering kernel and recovery of the eikonal coherence}
\label{sec:coulomb_kernel}

For the OCP correlator~\eqref{eq:K3},
\begin{equation}
\widetilde B(Q)
=
\widetilde K_3(Q,0)
=
\frac{4\pi C\kappa^2}{Q^2(Q^2+\kappa^2)}.
\label{eq:Btilde_Coulomb}
\end{equation}
The Wigner equation therefore becomes
\begin{align}
\partial_z I
+
\frac{\mathbf p}{k}\cdot\nabla_R I
&=
4\pi A_{\rm rel}^2C\kappa^2
\int\frac{d^2Q}{(2\pi)^2}
\nonumber\\
&\quad\times
\frac{
I(\mathbf R,\mathbf p-\mathbf Q,z)
-
I(\mathbf R,\mathbf p,z)
}{
Q^2(Q^2+\kappa^2)
}.
\label{eq:Wigner_Coulomb}
\end{align}
For small transverse momentum transfer $Q\ll\kappa$,
\begin{equation}
w(Q)\simeq\frac{4\pi A_{\rm rel}^2C}{Q^2}.
\label{eq:w_small_Q}
\end{equation}
The forward singularity is not a pathology because the collision operator contains the difference $I(\mathbf p-\mathbf Q)-I(\mathbf p)$.

If the beam is statistically homogeneous in $\mathbf R$, the streaming term vanishes. 
Transforming back to $\mathbf r$ gives
\begin{equation}
\partial_z\Gamma(\mathbf r,z)
=
-A_{\rm rel}^2[B(0)-B(\mathbf r)]\Gamma(\mathbf r,z).
\label{eq:Gamma_hom}
\end{equation}
For a fully coherent incident plane wave, $\Gamma(\mathbf r,0)=1$, hence
\begin{equation}
\Gamma(\mathbf r,L)
=
\exp[-A_{\rm rel}^2L(B(0)-B(\mathbf r))].
\label{eq:Gamma_solution}
\end{equation}
The remaining task is to evaluate $B(0)-B(r)$.

Using Eq.~\eqref{eq:Btilde_Coulomb},
\begin{equation}
B(0)-B(r)
=
\int\frac{d^2Q}{(2\pi)^2}
[1-e^{i\mathbf Q\cdot\mathbf r}]
\frac{4\pi C\kappa^2}{Q^2(Q^2+\kappa^2)}.
\label{eq:Bdiff_integral}
\end{equation}
By angular integration,
\begin{equation}
B(0)-B(r)
=
2C\kappa^2
\int_0^\infty dQ\,
\frac{1-J_0(Qr)}{Q(Q^2+\kappa^2)}.
\label{eq:Bdiff_J0}
\end{equation}
This integral has the closed form
\begin{equation}
B(0)-B(r)
=
2C
\left[
\ln(\kappa r)
+
K_0(\kappa r)
+
\gamma_E-\ln2
\right].
\label{eq:Bdiff_closed}
\end{equation}
For $\kappa r\ll1$,
\begin{equation}
B(0)-B(r)
\simeq
\frac{C(\kappa r)^2}{2}
\left[
\ln\frac{2}{\kappa r}+1-\gamma_E
\right],
\label{eq:Bdiff_small}
\end{equation}
while for $\kappa r\gg1$,
\begin{equation}
B(0)-B(r)
\simeq
2C[\ln(\kappa r)+\gamma_E-\ln2].
\label{eq:Bdiff_large}
\end{equation}
Substitution into Eq.~\eqref{eq:Gamma_solution} gives exactly the eikonal coherence function:
\begin{equation}
\Gamma(r,L)
=
\exp[-A_{\rm rel}^2L(B(0)-B(r))].
\label{eq:Gamma_eikonal_match}
\end{equation}
At large transverse separations,
\begin{equation}
\Gamma(r,L)
\propto
(\kappa r)^{-2A_{\rm rel}^2CL}.
\label{eq:Gamma_power}
\end{equation}
Thus the algebraic coherence tail is not tied to the straight-line picture alone. 
It is the homogeneous limit of the Bethe--Salpeter equation with the Coulomb Wigner kernel.

The Wigner representation also gives a controlled description of angular diffusion. 
For sufficiently small momentum transfers, expand the collision integral in $\mathbf Q$:
\begin{equation}
I(\mathbf p-\mathbf Q)-I(\mathbf p)
=
-Q_i\partial_{p_i}I
+
\frac{1}{2}Q_iQ_j\partial_{p_i}\partial_{p_j}I
+\cdots.
\label{eq:FP_expand}
\end{equation}
The linear term vanishes by isotropy. 
The second-order term yields the Fokker--Planck equation
\begin{equation}
\partial_z I
+
\frac{\mathbf p}{k}\cdot\nabla_R I
=
D_p\nabla_p^2 I,
\label{eq:FP}
\end{equation}
with
\begin{equation}
D_p
=
\frac{1}{4}
\int^{\Lambda}\frac{d^2Q}{(2\pi)^2}
w(Q)Q^2.
\label{eq:Dp_def}
\end{equation}
For the Coulomb kernel,
\begin{equation}
D_p(\Lambda)
=
\frac{A_{\rm rel}^2C\kappa^2}{4}
\ln\left(1+\frac{\Lambda^2}{\kappa^2}\right).
\label{eq:Dp}
\end{equation}
The logarithm in Eq.~\eqref{eq:Dp} is the Coulomb logarithm of the present
paraxial scattering problem. For $\Lambda\gg\kappa$,
\begin{equation}
\ln\left(1+\frac{\Lambda^2}{\kappa^2}\right)
\simeq
2\ln\frac{\Lambda}{\kappa}
=
2\ln\frac{\lambda_D}{a_{\rm micro}},
\end{equation}
where $a_{\rm micro}\sim\Lambda^{-1}$ is the microscopic ultraviolet cutoff
of the continuum Coulomb description. Thus the logarithm is generated by
the broad interval of transverse scales
$a_{\rm micro}\ll b\ll\lambda_D$, exactly as in the usual Coulomb
logarithm of plasma kinetic theory~\cite{pitaevskii2012physical}. The difference is that here it appears
in the Wigner angular-diffusion coefficient of a transmitted quantum wave,
rather than in a binary-collision transport coefficient.
The moment calculation for Eq.~\eqref{eq:FP} is given in Appendix~\ref{app:wigner_moments}. It yields
\begin{equation}
\langle p^2(z)\rangle=4D_pz,
\label{eq:p2}
\end{equation}
and, for an initially collimated beam,
\begin{equation}
\langle R^2(z)\rangle=\frac{4D_p}{3k^2}z^3.
\label{eq:R2}
\end{equation}
Thus, beyond eikonal dephasing, the same Coulomb disorder produces multiple small-angle scattering and transverse beam broadening.

\section{Physical implications}
\label{sec:implications}

The theory separates three effects. 
The first is attenuation of the disorder-averaged coherent amplitude, characterized by $\ell_{\rm rel}$. 
The second is loss of mutual coherence, described by $\Gamma(r,L)$ or $\gamma(r)$. 
The third is redistribution of intensity in transverse phase space, described by the Wigner kinetic equation. 
Only the last of these is a conservative transport process. 
The first two are coherence losses after disorder averaging.

The relativistic saturation of $A_{\rm rel}$ affects all three. 
The coherence exponent
\begin{equation}
\eta_{\rm rel}=2A_{\rm rel}^2CL
\label{eq:eta_rel}
\end{equation}
has a lower high-energy bound
\begin{equation}
\eta_{\rm min}=\frac{2CL}{\hbar^2c^2}.
\label{eq:eta_min}
\end{equation}
Similarly, the eikonal attenuation length has the upper bound~\eqref{eq:ell_max}, and the angular diffusion coefficient has the high-energy floor
\begin{equation}
D_{p,\min}
=
\frac{C\kappa^2}{4\hbar^2c^2}
\ln\left(1+\frac{\Lambda^2}{\kappa^2}\right).
\label{eq:Dp_min}
\end{equation}
Therefore, higher voltage cannot remove Coulomb-induced phase noise; it can only bring the system closer to the saturated limit.
A useful way to quantify this approach is through the ratio
\begin{equation}
\frac{A_{\rm rel}}{A_\infty}=\frac{c}{v}=\frac{1}{\beta},
\qquad
A_\infty=\frac{1}{\hbar c}.
\label{eq:A_ratio}
\end{equation}
For representative electron kinetic energies, $K=80$, $200$, and $300~{\rm keV}$, one has $\beta\simeq0.50$, $0.70$, and $0.78$, respectively. Thus $A_{\rm rel}/A_\infty\simeq1.99$, $1.44$, and $1.29$. Standard TEM energies therefore do not remove Coulomb-induced phase noise; rather, they already lie in the crossover toward the relativistic saturation regime in which further increases of beam energy produce progressively smaller reductions of the Coulomb dephasing strength.

For liquid-cell TEM this statement should not be confused with an instrumental resolution bound. 
The electron wavelength, lens aberrations, detector response, radiation damage, and inelastic scattering remain separate constraints. 
The present result concerns the stochastic phase and coherence degradation induced by thermal Coulomb fluctuations of the sample. 
In this sense the Coulomb-disorder scale plays a role analogous to a statistical wave-front limit, in the spirit of coherence limits such as the Fried parameter in atmospheric optics~\cite{Fried1966}: it is set by the medium and the propagation length, not by the objective lens alone.

The Bethe--Salpeter derivation also clarifies the status of multiple scattering. 
The eikonal approximation is controlled when transverse deflections are negligible. 
The Wigner equation, by contrast, remains a kinetic description of accumulated small-angle scattering. 
In a homogeneous system it reduces exactly to the same mutual-coherence decay because the streaming term is absent. 
For finite beams, apertures, or inhomogeneous illumination, the streaming term couples coherence loss to angular diffusion and beam broadening. This is the natural framework for comparing the present theory with simulations or experiments in realistic microscope geometries.

Finally, the logarithmic long-distance form of $\Gamma(r,L)$ suggests an infrared interpretation in terms of an effective two-dimensional Gaussian phase field. 
Such an interpretation may be useful for discussing possible topological defects of the transmitted wave front. 
However, the present paper does not rely on a topological transition. 
The robust and directly testable conclusions are the Coulomb-induced algebraic mutual decoherence, the kinetic small-angle scattering kernel, and the relativistic saturation of the effective coupling.

\section{Conclusion}
\label{sec:conclusion}

We have formulated a relativistic paraxial theory of Coulomb-induced decoherence of electron matter waves in a Coulomb-fluctuating medium. 
The effective coupling of the electron envelope to the random electrostatic potential is
\begin{equation}
A_{\rm rel}=\frac{1}{\hbar v},
\end{equation}
which saturates at $1/(\hbar c)$ in the ultra-relativistic limit. 
This saturation implies a finite high-energy floor for Coulomb phase noise and therefore explains why increasing the TEM voltage yields diminishing returns for suppressing Coulomb-disorder-induced coherence loss.

In the eikonal approximation, the electron accumulates a random phase along a straight path. 
The OCP potential correlator produces a logarithmic phase structure function and an algebraic decay of the mutual coherence function. 
The coherence length obeys $\rho_c\sim\lambda_D\sqrt{\ell_{\rm rel}/L}$, with a relativistically saturated attenuation scale.

We then derived the Bethe--Salpeter equation for the two-point coherence propagator directly from the paraxial equation. 
In the Markov limit the equation contains a dephasing kernel $A_{\rm rel}^2[B(0)-B(r)]$. 
After Wigner transformation it becomes a kinetic equation with the Coulomb scattering kernel
\begin{equation}
w(Q)=
\frac{4\pi A_{\rm rel}^2C\kappa^2}{Q^2(Q^2+\kappa^2)}.
\end{equation}
This kinetic equation describes conservative angular diffusion and transverse broadening, while its homogeneous coordinate-space limit reproduces the eikonal mutual coherence exactly.

The resulting framework provides a bridge between electron coherence theory, wave propagation in random media, and kinetic multiple-scattering theory. 
It also gives a practical message for liquid-cell and cryogenic electron microscopy: Coulomb disorder in the sample can impose a medium-induced electron-wave coherence limit whose suppression saturates at relativistic beam energies.

\section*{Acknowledgments}
The research leading to these results has received funding from the Basic Research Program at HSE University (HSE-BR-2025-007).

\appendix

\appendix

\section{Potential-energy correlator in a one-component plasma}
\label{app:ocp_correlator}

In this appendix we derive the static correlation function of the
electrostatic potential energy experienced by a charged particle
embedded in a weakly coupled one-component plasma (OCP). The result is
used in the main text as the input correlator of the random Coulomb
potential.

We consider an OCP consisting of mobile particles of charge \(q_0\) and
mean number density \(n_0\), embedded in a uniform neutralizing
background. The dielectric constant of the medium is denoted by
\(\epsilon\), and we use the Coulomb convention
\begin{equation}
  v_{00}(r)=\frac{q_0^2}{\epsilon r},
  \qquad
  \widetilde v_{00}(k)=\frac{4\pi q_0^2}{\epsilon k^2}.
  \label{eq:app_v00}
\end{equation}
The neutralizing background removes the zero Fourier mode. All
expressions below are therefore understood for \(k\neq 0\), with the
long-wavelength limit taken after imposing global charge neutrality.

Let
\begin{equation}
  \delta n({\bf r})=n({\bf r})-n_0
\end{equation}
be the fluctuation of the microscopic number density. We use the
Fourier convention
\begin{equation}
  \delta n_{\bf k}
  =
  \int d^3 r\, e^{-i{\bf k}\cdot{\bf r}}\delta n({\bf r}),
  \qquad
  \delta n({\bf r})
  =
  \int\frac{d^3 k}{(2\pi)^3}\,
  e^{i{\bf k}\cdot{\bf r}}\delta n_{\bf k}.
  \label{eq:app_fourier}
\end{equation}
The static structure factor is defined by
\begin{equation}
  \left\langle
  \delta n_{\bf k}\delta n_{\bf k'}
  \right\rangle
  =
  (2\pi)^3\delta({\bf k}+{\bf k}')\,n_0 S(k).
  \label{eq:app_S_def}
\end{equation}

Within the Debye--Huckel, or equivalently random-phase, approximation
the OCP structure factor is
\begin{equation}
  S(k)
  =
  \frac{1}{1+\beta n_0\widetilde v_{00}(k)}
  =
  \frac{k^2}{k^2+\kappa^2},
  \label{eq:app_S_RPA}
\end{equation}
where
\begin{equation}
  \beta=\frac{1}{k_B T},
  \qquad
  \kappa^2=\frac{4\pi\beta q_0^2 n_0}{\epsilon}
  \label{eq:app_kappa}
\end{equation}
is the inverse Debye length squared. Equation
\eqref{eq:app_S_RPA} also displays explicitly the Stillinger--Lovett
small-\(k\) behavior \(S(k)\sim k^2/\kappa^2\), which ensures perfect
screening of charge fluctuations.

Now consider a probe particle of charge \(q_{\rm p}\). Its electrostatic
potential energy due to the OCP density fluctuation is
\begin{equation}
  W({\bf r})
  =
  \int d^3 r'\,
  \frac{q_{\rm p}q_0}{\epsilon|{\bf r}-{\bf r}'|}
  \delta n({\bf r}').
  \label{eq:app_W_real}
\end{equation}
In Fourier representation,
\begin{equation}
  W_{\bf k}
  =
  \widetilde v_{\rm p0}(k)\delta n_{\bf k},
  \qquad
  \widetilde v_{\rm p0}(k)
  =
  \frac{4\pi q_{\rm p}q_0}{\epsilon k^2}.
  \label{eq:app_W_fourier}
\end{equation}
The potential-energy correlator is defined by
\begin{equation}
  K_W({\bf r})
  =
  \left\langle W({\bf r})W({\bf 0})\right\rangle
  =
  \int\frac{d^3 k}{(2\pi)^3}\,
  e^{i{\bf k}\cdot{\bf r}}\widetilde K_W(k),
  \label{eq:app_KW_def}
\end{equation}
where
\begin{equation}
  \left\langle
  W_{\bf k}W_{\bf k'}
  \right\rangle
  =
  (2\pi)^3\delta({\bf k}+{\bf k}')\widetilde K_W(k).
  \label{eq:app_KW_spectrum_def}
\end{equation}
Using Eqs.~\eqref{eq:app_S_def} and \eqref{eq:app_W_fourier}, we obtain
\begin{equation}
  \widetilde K_W(k)
  =
  \widetilde v_{\rm p0}^{\,2}(k)\, n_0 S(k).
  \label{eq:app_KW_spectrum_1}
\end{equation}
Substitution of the RPA structure factor gives
\begin{align}
  \widetilde K_W(k)
  &=
  \left(
  \frac{4\pi q_{\rm p}q_0}{\epsilon k^2}
  \right)^2
  n_0
  \frac{k^2}{k^2+\kappa^2}
  \nonumber\\
  &=
  \frac{16\pi^2 q_{\rm p}^2q_0^2 n_0}{\epsilon^2}
  \frac{1}{k^2(k^2+\kappa^2)}.
  \label{eq:app_KW_spectrum_2}
\end{align}
Using the definition \eqref{eq:app_kappa}, this can be written as
\begin{equation}
  \widetilde K_W(k)
  =
  4\pi C_{\rm p}
  \frac{\kappa^2}{k^2(k^2+\kappa^2)},
  \qquad
  C_{\rm p}
  =
  \frac{k_B T q_{\rm p}^2}{\epsilon}.
  \label{eq:app_KW_spectrum_final}
\end{equation}
Equivalently,
\begin{equation}
  \widetilde K_W(k)
  =
  4\pi C_{\rm p}
  \left(
  \frac{1}{k^2}
  -
  \frac{1}{k^2+\kappa^2}
  \right).
  \label{eq:app_KW_decomposition}
\end{equation}
Using the standard Fourier transforms
\begin{equation}
  \int\frac{d^3 k}{(2\pi)^3}\,
  e^{i{\bf k}\cdot{\bf r}}
  \frac{4\pi}{k^2}
  =
  \frac{1}{r},
  \qquad
  \int\frac{d^3 k}{(2\pi)^3}\,
  e^{i{\bf k}\cdot{\bf r}}
  \frac{4\pi}{k^2+\kappa^2}
  =
  \frac{e^{-\kappa r}}{r},
  \label{eq:app_fourier_transforms}
\end{equation}
we finally obtain
\begin{equation}
  K_W(r)
  =
  \left\langle W({\bf r})W({\bf 0})\right\rangle
  =
  C_{\rm p}
  \frac{1-e^{-\kappa r}}{r},
  \qquad
  C_{\rm p}
  =
  \frac{k_B T q_{\rm p}^2}{\epsilon}.
  \label{eq:app_KW_real_final}
\end{equation}
For a particle belonging to the same OCP species, \(q_{\rm p}=q_0\),
and therefore
\begin{equation}
  C_{\rm p}\equiv C
  =
  \frac{k_B T q_0^2}{\epsilon}.
  \label{eq:app_C_same_species}
\end{equation}

Several remarks are useful. First, the correlator is finite at the
origin:
\begin{equation}
  K_W(0)
  =
  \lim_{r\to 0}
  C_{\rm p}\frac{1-e^{-\kappa r}}{r}
  =
  C_{\rm p}\kappa .
  \label{eq:app_KW_zero}
\end{equation}
Second, at distances large compared with the Debye length,
\begin{equation}
  K_W(r)
  \simeq
  \frac{C_{\rm p}}{r},
  \qquad
  r\gg \kappa^{-1}.
  \label{eq:app_KW_large_r}
\end{equation}
This long-range \(1/r\) tail is not in contradiction with charge
screening. It reflects the fact that the potential is obtained from the
charge density by an inverse Laplacian. The screened small-\(k\)
behavior of the density structure factor, \(S(k)\propto k^2\), is
converted into \(\widetilde K_W(k)\propto 1/k^2\), and therefore into a
Coulombic potential-energy correlation in real space.

The eikonal theory developed in the main text involves the longitudinal
projection of this three-dimensional correlator. It is convenient to
define
\begin{equation}
  B({\boldsymbol\rho})
  =
  \int_{-\infty}^{\infty} dz\,
  K_W\!\left(\sqrt{\rho^2+z^2}\right).
  \label{eq:app_B_def}
\end{equation}
The quantity \(B({\boldsymbol\rho})\) itself contains an additive
infrared-divergent constant, but the difference entering the phase
structure function is finite:
\begin{align}
  B({\bf 0})-B({\boldsymbol\rho})
  &=
  \int\frac{d^2 Q}{(2\pi)^2}
  \left[
  1-e^{i{\bf Q}\cdot{\boldsymbol\rho}}
  \right]
  \widetilde K_W(Q,k_z=0)
  \nonumber\\
  &=
  2C_{\rm p}\kappa^2
  \int_0^\infty dQ\,
  \frac{1-J_0(Q\rho)}{Q(Q^2+\kappa^2)}
  \nonumber\\
  &=
  2C_{\rm p}
  \left[
  \ln\left(\frac{\kappa\rho}{2}\right)
  +\gamma_E
  +K_0(\kappa\rho)
  \right],
  \label{eq:app_B_difference}
\end{align}
where \(J_0\) is the Bessel function, \(K_0\) is the modified Bessel
function of the second kind, and \(\gamma_E\) is Euler's constant. In
particular,
\begin{equation}
  B({\bf 0})-B({\boldsymbol\rho})
  \simeq
  \frac{C_{\rm p}(\kappa\rho)^2}{2}
  \left[
  \ln\left(\frac{2}{\kappa\rho}\right)
  +1-\gamma_E
  \right],
  \qquad
  \kappa\rho\ll 1,
  \label{eq:app_B_small_rho}
\end{equation}
whereas
\begin{equation}
  B({\bf 0})-B({\boldsymbol\rho})
  \simeq
  2C_{\rm p}
  \left[
  \ln(\kappa\rho)+\gamma_E-\ln 2
  \right],
  \qquad
  \kappa\rho\gg 1.
  \label{eq:app_B_large_rho}
\end{equation}
These formulas are the static OCP input used in the eikonal and
Bethe--Salpeter parts of the theory.
\section{Paraxial reduction of the Dirac equation}
\label{app:dirac}

Here we give a self-contained derivation of the relativistic paraxial equation used in the main text. The external field is a weak static scalar potential energy $W(\mathbf r)$. We retain all terms linear in $W$ in the leading scalar part of the Dirac reduction and neglect only gradient corrections generated by spatial derivatives of $W$, such as Darwin and spin-orbit-like terms, which are of higher paraxial order for the present phase-noise problem.

For a static scalar potential energy $W(\mathbf r)$, the stationary Dirac equation is
\begin{equation}
\left[c\bm\alpha\cdot\hat{\mathbf p}+\beta mc^2+W(\mathbf r)\right]\Psi=E\Psi .
\label{eq:Dirac_app}
\end{equation}
Writing the bispinor as
\begin{equation}
\Psi=
\begin{pmatrix}
\varphi\\
\eta
\end{pmatrix},
\end{equation}
one obtains
\begin{equation}
(E-W-mc^2)\varphi
=
c\bm\sigma\cdot\hat{\mathbf p}\,\eta,
\label{eq:dirac_upper_comp_app}
\end{equation}
\begin{equation}
(E-W+mc^2)\eta
=
c\bm\sigma\cdot\hat{\mathbf p}\,\varphi .
\label{eq:dirac_lower_comp_app}
\end{equation}
Thus
\begin{equation}
\eta
=
\frac{c\bm\sigma\cdot\hat{\mathbf p}}
{E-W+mc^2}\,\varphi .
\label{eq:eta_elim_app}
\end{equation}
Substitution into Eq.~\eqref{eq:dirac_upper_comp_app} gives
\begin{equation}
(E-W-mc^2)\varphi
=
c\bm\sigma\cdot\hat{\mathbf p}
\left[
\frac{c\bm\sigma\cdot\hat{\mathbf p}}
{E-W+mc^2}\varphi
\right].
\label{eq:upper_exact_elim_app}
\end{equation}
In the leading paraxial and weak-field approximation we neglect gradients of $W$ in the denominator. These gradients generate higher-order Darwin and spin-orbit-like terms which are not needed in the present scalar phase-noise problem. However, the value of $W$ itself must be kept in the denominator when the result is expanded to first order in $W$. With this convention,
\begin{equation}
(E-W-mc^2)(E-W+mc^2)\varphi
\simeq
c^2(\bm\sigma\cdot\hat{\mathbf p})^2\varphi .
\label{eq:kg_like_step_app}
\end{equation}
In the absence of a magnetic field,
\begin{equation}
(\bm\sigma\cdot\hat{\mathbf p})^2=\hat{\mathbf p}^{\,2},
\end{equation}
and therefore
\begin{equation}
\left[(E-W)^2-m^2c^4\right]\varphi
\simeq
c^2\hat{\mathbf p}^{\,2}\varphi .
\label{eq:kg_like_app}
\end{equation}
Since $\hat{\mathbf p}^{\,2}=-\hbar^2\nabla^2$, this is equivalent to
\begin{equation}
\nabla^2\varphi
+
\frac{(E-W)^2-m^2c^4}{\hbar^2c^2}\varphi=0 .
\label{eq:scalar_wave_app}
\end{equation}

We now expand the coefficient of $\varphi$ to first order in the potential:
\begin{equation}
(E-W)^2-m^2c^4
=
E^2-m^2c^4-2EW+O(W^2).
\label{eq:linear_W_app}
\end{equation}
For the unperturbed relativistic particle,
\begin{equation}
E=\gamma mc^2,
\qquad
p_0=\gamma mv,
\qquad
E^2-m^2c^4=p_0^2c^2 .
\end{equation}
Introducing
\begin{equation}
k=\frac{p_0}{\hbar}=\frac{\gamma mv}{\hbar},
\label{eq:k_def_app}
\end{equation}
Eq.~\eqref{eq:scalar_wave_app} becomes
\begin{equation}
\nabla^2\varphi
+
\left[
k^2-
\frac{2E}{\hbar^2c^2}W(\mathbf r)
\right]\varphi=0 .
\label{eq:scalar_linear_app}
\end{equation}
We introduce the slowly varying paraxial envelope
\begin{equation}
\varphi(\bm\rho,z)=e^{ikz}\psi(\bm\rho,z).
\label{eq:ansatz_app}
\end{equation}
Then
\begin{equation}
\nabla^2\varphi
=
e^{ikz}
\left(
\nabla_\perp^2+
\partial_z^2+
2ik\partial_z-k^2
\right)\psi .
\label{eq:laplace_envelope_app}
\end{equation}
Substituting this into Eq.~\eqref{eq:scalar_linear_app} gives
\begin{equation}
\left[
\nabla_\perp^2+
\partial_z^2+
2ik\partial_z-
\frac{2E}{\hbar^2c^2}W
\right]\psi=0 .
\label{eq:before_paraxial_app}
\end{equation}
Neglecting $\partial_z^2\psi$ compared with $k\partial_z\psi$, we obtain
\begin{equation}
2ik\partial_z\psi
=
-\nabla_\perp^2\psi
+
\frac{2E}{\hbar^2c^2}W\psi .
\label{eq:paraxial_step_app}
\end{equation}
Therefore
\begin{equation}
i\partial_z\psi
=
-\frac{1}{2k}\nabla_\perp^2\psi
+
\frac{E}{\hbar^2c^2k}W\psi .
\label{eq:paraxial_coeff_step_app}
\end{equation}
Using $E=\gamma mc^2$ and $k=\gamma mv/\hbar$, the coefficient of the scalar potential is
\begin{equation}
\frac{E}{\hbar^2c^2k}
=
\frac{\gamma mc^2}{\hbar^2c^2(\gamma mv/\hbar)}
=
\frac{1}{\hbar v} .
\label{eq:Arel_derivation_app}
\end{equation}
Thus the leading paraxial equation is
\begin{equation}
i\partial_z\psi
=
\left[
-\frac{1}{2k}\nabla_\perp^2
+
\frac{1}{\hbar v}W(\bm\rho,z)
\right]\psi .
\label{eq:paraxial_app}
\end{equation}
This is Eq.~\eqref{eq:paraxial} of the main text, with
\begin{equation}
A_{\rm rel}=\frac{1}{\hbar v}.
\label{eq:Arel_final_app}
\end{equation}
The same coefficient follows immediately from the eikonal Hamilton--Jacobi argument. At fixed total energy,
\begin{equation}
(E-W)^2=m^2c^4+p^2c^2 .
\label{eq:HJ_dispersion_app}
\end{equation}
The momentum in the weak potential is
\begin{equation}
p(W)=\frac{1}{c}\sqrt{(E-W)^2-m^2c^4} .
\end{equation}
Hence, to first order in $W$,
\begin{equation}
\delta p
=
-\frac{E}{c^2p_0}W
=
-\frac{W}{v},
\label{eq:dp_eik_app}
\end{equation}
where $v=p_0c^2/E$. The corresponding eikonal phase shift is therefore
\begin{equation}
\delta\phi
=
\frac{1}{\hbar}\int dz\,\delta p
=
-\frac{1}{\hbar v}\int dz\,W .
\label{eq:eikonal_phase_app}
\end{equation}
This confirms the paraxial result.

The coefficient therefore saturates solely because the particle velocity is bounded by the speed of light:
\begin{equation}
A_{\rm rel}=\frac{1}{\hbar v}
=\frac{1}{\hbar c\beta}
\longrightarrow
\frac{1}{\hbar c},
\qquad \beta=\frac{v}{c},\quad \gamma\to\infty .
\label{eq:A_saturation_app}
\end{equation}

\section{Evaluation of the Coulomb dephasing kernel}
\label{app:B_kernel}

The longitudinally integrated transverse correlator has Fourier representation
\begin{equation}
\widetilde B(Q)=\widetilde K_3(Q,0).
\end{equation}
Using Eq.~\eqref{eq:K3},
\begin{equation}
\widetilde B(Q)
=
\frac{4\pi C\kappa^2}{Q^2(Q^2+\kappa^2)}.
\end{equation}
The difference entering the Bethe--Salpeter equation is
\begin{equation}
B(0)-B(r)
=
2C\kappa^2
\int_0^\infty dQ\,
\frac{1-J_0(Qr)}{Q(Q^2+\kappa^2)}.
\end{equation}
Introducing $x=\kappa r$, differentiating with respect to $x$, and using the standard Bessel transform
\begin{equation}
\frac{d}{dx}[B(0)-B(r)]
=
2C\left[\frac{1}{x}-K_1(x)\right],
\end{equation}
one obtains
\begin{equation}
B(0)-B(r)
=
2C[\ln x+K_0(x)+\gamma_E-\ln2],
\qquad x=\kappa r.
\end{equation}
The integration constant is fixed by the condition $B(0)-B(0)=0$.

\section{Moment equations in the Wigner--Fokker--Planck limit}
\label{app:wigner_moments}

In this Appendix we derive the moment relations used in the main text for the Wigner--Fokker--Planck approximation. The starting point is
\begin{equation}
\partial_z I(\mathbf R,\mathbf p,z)
+
\frac{\mathbf p}{k}\cdot\nabla_{\mathbf R} I(\mathbf R,\mathbf p,z)
=
D_p\nabla_{\mathbf p}^2 I(\mathbf R,\mathbf p,z),
\label{eq:FP_mom_app}
\end{equation}
where \(\mathbf R\) is the transverse coordinate, \(\mathbf p\) is the transverse wave vector, and \(D_p\) is the angular-diffusion coefficient. The derivation below does not depend on the microscopic origin of \(D_p\).

We define phase-space averages by
\begin{equation}
\langle F\rangle
=
\frac{1}{N}
\int d^2R
\int\frac{d^2p}{(2\pi)^2}\,
F(\mathbf R,\mathbf p)
I(\mathbf R,\mathbf p,z),
\label{eq:average_mom_app}
\end{equation}
where
\begin{equation}
N=
\int d^2R
\int\frac{d^2p}{(2\pi)^2}\,
I(\mathbf R,\mathbf p,z).
\label{eq:norm_mom_app}
\end{equation}
Equation~\eqref{eq:FP_mom_app} conserves \(N\), provided that the Wigner function is sufficiently localized in \(\mathbf R\) and \(\mathbf p\), or that the boundary terms vanish. This assumption will be used throughout the integration by parts below.

We first consider the transverse momentum variance,
\begin{equation}
p^2=p_x^2+p_y^2.
\end{equation}
Multiplying Eq.~\eqref{eq:FP_mom_app} by \(p^2\) and integrating over phase space gives
\begin{equation}
\begin{aligned}
\frac{d}{dz}\langle p^2\rangle
=&
-\frac{1}{N}
\int d^2R
\int\frac{d^2p}{(2\pi)^2}\,
p^2\,
\frac{\mathbf p}{k}\cdot\nabla_{\mathbf R}I
\\
&+
\frac{D_p}{N}
\int d^2R
\int\frac{d^2p}{(2\pi)^2}\,
p^2\nabla_{\mathbf p}^2I .
\end{aligned}
\label{eq:p2_derivation_mom_app}
\end{equation}
The first term vanishes after integration by parts with respect to \(\mathbf R\), since \(p^2\) is independent of \(\mathbf R\). For the second term, integration by parts twice in \(\mathbf p\) gives
\begin{equation}
\int\frac{d^2p}{(2\pi)^2}\,
p^2\nabla_{\mathbf p}^2I
=
\int\frac{d^2p}{(2\pi)^2}\,
I\nabla_{\mathbf p}^2p^2 .
\label{eq:p2_ibp_mom_app}
\end{equation}
In two transverse dimensions,
\begin{equation}
\nabla_{\mathbf p}^2p^2
=
\left(
\frac{\partial^2}{\partial p_x^2}
+
\frac{\partial^2}{\partial p_y^2}
\right)
(p_x^2+p_y^2)
=
4.
\label{eq:lap_p2_mom_app}
\end{equation}
Therefore,
\begin{equation}
\frac{d}{dz}\langle p^2\rangle
=
4D_p .
\label{eq:p2_derivative_mom_app}
\end{equation}
For an initially collimated beam, \(\langle p^2(0)\rangle=0\), this gives
\begin{equation}
\langle p^2(z)\rangle
=
4D_p z .
\label{eq:p2_result_mom_app}
\end{equation}
This is Eq.~\eqref{eq:p2} of the main text.

We now derive the transverse spatial broadening. Let
\begin{equation}
R^2=R_x^2+R_y^2.
\end{equation}
Multiplying Eq.~\eqref{eq:FP_mom_app} by \(R^2\) and integrating over phase space gives
\begin{equation}
\begin{aligned}
\frac{d}{dz}\langle R^2\rangle
=&
-\frac{1}{N}
\int d^2R
\int\frac{d^2p}{(2\pi)^2}\,
R^2
\frac{\mathbf p}{k}\cdot\nabla_{\mathbf R}I
\\
&+
\frac{D_p}{N}
\int d^2R
\int\frac{d^2p}{(2\pi)^2}\,
R^2\nabla_{\mathbf p}^2I .
\end{aligned}
\label{eq:R2_derivation_mom_app}
\end{equation}
The diffusion term vanishes because \(R^2\) is independent of \(\mathbf p\). Integrating the remaining streaming term by parts in \(\mathbf R\), we obtain
\begin{equation}
\frac{d}{dz}\langle R^2\rangle
=
\frac{1}{k}
\left\langle
\mathbf p\cdot\nabla_{\mathbf R}R^2
\right\rangle .
\label{eq:R2_streaming_mom_app}
\end{equation}
Since
\begin{equation}
\nabla_{\mathbf R}R^2=2\mathbf R,
\label{eq:grad_R2_mom_app}
\end{equation}
one finds
\begin{equation}
\frac{d}{dz}\langle R^2\rangle
=
\frac{2}{k}
\langle \mathbf R\cdot\mathbf p\rangle .
\label{eq:R2_mixed_mom_app}
\end{equation}

It remains to determine the mixed moment \(\langle \mathbf R\cdot\mathbf p\rangle\). Multiplying Eq.~\eqref{eq:FP_mom_app} by \(\mathbf R\cdot\mathbf p\) and integrating over phase space gives
\begin{equation}
\begin{aligned}
\frac{d}{dz}
\langle \mathbf R\cdot\mathbf p\rangle
=&
-\frac{1}{N}
\int d^2R
\int\frac{d^2p}{(2\pi)^2}\,
(\mathbf R\cdot\mathbf p)
\frac{\mathbf p}{k}\cdot\nabla_{\mathbf R}I
\\
&+
\frac{D_p}{N}
\int d^2R
\int\frac{d^2p}{(2\pi)^2}\,
(\mathbf R\cdot\mathbf p)
\nabla_{\mathbf p}^2I .
\end{aligned}
\label{eq:Rp_derivation_mom_app}
\end{equation}
The diffusion term vanishes after integration by parts in \(\mathbf p\), because \(\mathbf R\cdot\mathbf p\) is linear in \(\mathbf p\):
\begin{equation}
\nabla_{\mathbf p}^2(\mathbf R\cdot\mathbf p)=0.
\label{eq:lap_Rp_mom_app}
\end{equation}
The streaming term gives
\begin{equation}
\frac{d}{dz}
\langle \mathbf R\cdot\mathbf p\rangle
=
\frac{1}{k}
\left\langle
\mathbf p\cdot\nabla_{\mathbf R}
(\mathbf R\cdot\mathbf p)
\right\rangle .
\label{eq:Rp_streaming_mom_app}
\end{equation}
Since
\begin{equation}
\nabla_{\mathbf R}
(\mathbf R\cdot\mathbf p)
=
\mathbf p,
\label{eq:grad_Rp_mom_app}
\end{equation}
we obtain
\begin{equation}
\frac{d}{dz}
\langle \mathbf R\cdot\mathbf p\rangle
=
\frac{1}{k}\langle p^2\rangle .
\label{eq:Rp_derivative_mom_app}
\end{equation}
Using Eq.~\eqref{eq:p2_result_mom_app}, and assuming that the initial beam has no coordinate--momentum correlation,
\begin{equation}
\langle \mathbf R\cdot\mathbf p\rangle_{z=0}=0,
\end{equation}
we get
\begin{equation}
\begin{aligned}
\langle \mathbf R\cdot\mathbf p\rangle
&=
\frac{1}{k}
\int_0^z dz'\,\langle p^2(z')\rangle
\\
&=
\frac{1}{k}
\int_0^z dz'\,4D_p z'
=
\frac{2D_p}{k}z^2 .
\end{aligned}
\label{eq:Rp_result_mom_app}
\end{equation}
Substitution into Eq.~\eqref{eq:R2_mixed_mom_app} yields
\begin{equation}
\frac{d}{dz}\langle R^2\rangle
=
\frac{4D_p}{k^2}z^2 .
\label{eq:R2_growth_rate_mom_app}
\end{equation}
Finally, for an initially narrow transverse distribution, \(\langle R^2(0)\rangle=0\), integration gives
\begin{equation}
\langle R^2(z)\rangle
=
\frac{4D_p}{3k^2}z^3 .
\label{eq:R2_result_mom_app}
\end{equation}
This is Eq.~\eqref{eq:R2} of the main text.

Equations~\eqref{eq:p2_result_mom_app} and \eqref{eq:R2_result_mom_app} show that the transverse momentum variance grows linearly with propagation distance, while the transverse spatial variance grows as \(z^3\). This cubic law reflects the fact that the transverse coordinate is obtained by integrating the random angular deflection accumulated during propagation.

\bibliographystyle{iopart-num}
\bibliography{main_JPhysB_Coulomb_decoherence_checked}

\end{document}